\documentclass[aps,pra,twocolumn,superscriptaddress,showkeys,10pt]{revtex4-1}

\usepackage{graphicx}
\usepackage{xcolor}

\begin{document}


\title{Lithiation - delithiation cycles of amorphous Si nanowires investigated by molecular dynamics simulations}


\author{Julien Godet}
\email[]{julien.godet@univ-poitiers.fr}
\affiliation{
Department of Physics and Mechanics of Materials, CNRS, Institut Pprime, Universit\'e de Poitiers, TSA 41123, F-86073, Poitiers Cedex 9, France
}

\author{Teut\"e Bunjaku}

\author{Mathieu Luisier}
\affiliation{Integrated Systems Laboratory, Department of Electrical Engineering and Information Technology, ETH Zurich, Gloriastrasse 35, 8092 Zurich, Switzerland}

\date{\today}

\begin{abstract}

The atomistic mechanisms during lithiation and delithiation of amorphous Si nanowires ($a$-SiNW)  have been investigated over cycles by molecular dynamics simulations. First, the Modified Embedded Atom Method (MEAM) potential from Cui et al. [J. Power Sources. 2012, {\bf207} 150]  has been further optimized on static  (Li$_x$Si alloy phases and point defect energies) and on dynamic properties (Li diffusion) to reproduce the lithiation of small crystalline Si nanowires calculated at the {\it ab initio} level. The lithiation of  $a$-SiNW reveals a two-phase process of lithiation with a larger diffusion interface compared to crystalline Si lithiation.  Compressive axial stresses are observed in the amorphous Si$_x$Li alloy outer shell. They are easily released thanks to the soft glassy behavior of the amorphous alloy. Conversely, in crystalline SiNW, the larger stress in the narrow crystalline lithiated interface is hardly released and requires a phase transformation to amorphous to operate, which delays the lithiation. The history of the charge-discharge cycles as well as the temperature appear as driving forces for phase transformation from amorphous Li$_x$Si alloy to  the more stable crystalline phase counterpart. Our work suggest that a full delithiation could heal the SiNWs to improve the life cycles of Li-ion batteries with Si anode.

\end{abstract}

\keywords{Li-ion battery, stress, crystallization, Li diffusion, DFT, MEAM, potential optimization}

\maketitle

\section{Introduction}

In the race of high power portable Li-ion batteries, silicon is considered as one of the most promising candidates for anodes due to its very high theoretical capacity of 3579 mA h g$^{-1}$ \cite{Obrovac04ESSL}, almost 10 times larger than current graphite anodes. In addition, silicon is very abundant on earth  and the industries have a large experience for its manufacture. However, the main problems \cite{Ashuri16Nanoscale} that hindered practical application are low intrinsic electrical conductivity, instability of solid electrolyte interphase \cite{Cao17AMI}, and large volume modifications during lithiation / delithiation (300\%) that cause fractures and fading capacity \cite{Lee12PNAS,Liu12ACSNano}. Some of these issues can be partly solved by proper design considerations that allow for better life cycles of the battery cells \cite{Ashuri16Nanoscale}. In particular, the nanostructuration of the Si anode, as reviewed by  McDowell {\it et al.}  \cite{McDowell13AM}, results in better accommodation of the lithiation-induced strain and delays the anode fracture, highlighting the importance to   understand the mechanism of volume change in nanostructures and the link between Li diffusion, strain, plasticity, and  fracture.  

Experimentally, the stress evolution in silicon thin film has been monitored during lithiation and delithiation  \cite{Sethuraman10JPS}. It is related to the high volume change of Li$_x$Si alloys. 
In crystalline Si nanowire ($c$-SiNW), a two-phase lithiation mechanism composed of pristine Si in the core and  amorphous Li$_x$Si alloy ($x$=3.75) in the outer shell has been identified, with a sharp interface of 1 nm, where the lithiation operates by atomic layer peeling \cite{Liu12NatNano}.
The self-limiting lithiation of $c$-SiNW appears related to the stress generated during lithiation \cite{Liu13Nano}. Conversely to the first lithiation of crystalline silicon, the delithiation and the second lithiation are described as a single-phase process because of the amorphous structure of the anode after lithiation and/or delithiation  \cite{Cao17AMI}. However, the lithiation of amorphous Si nanospheres revealed a two-phase process  \cite{McDowell13NL}, raising  questions about the exact mechanism behind the lithiation of amorphous Si phases. The latter study also claims that the delithiation leads to nanospheres with a volume 25\% larger than before lithiation. These observations call for a deeper comprehension of the lithiation and delithiation at the atomic scale.

Atomistic simulations are attractive as a complementary approach to experiments, because they can access relatively easily to the underlying physical mechanisms, while providing a deep insight into experimental observations.  {\it Ab initio} calculations have  been performed in bulk systems to understand the structural transformation in lithiated crystalline \cite{Zhao11NL} and in amorphous silicon \cite{Cubuk14NL}, or the first stage of lithiation on model crystalline \cite{Liu12NatNano,Seo15RSCA}  and amorphous  \cite{Wang13NL}  Si interfaces.
Such a technique has also revealed a two-phase lithiation mechanism in amorphous Si nanoclusters \cite{Pedersen17PRA}, and $c$-SiNW \cite{Bunjaku} despite the limited size and time scale accessible.

To access to larger sizes and/or time scales, classical Molecular Dynamics (MD) simulations can be helpful although their results depend on the quality of the empirical potential used. Two classes of potential are available for Li-Si interactions. ReaxFF potentials have been used for the modeling of lithiation and delithiation of $c$-SiNW and to monitor the related stress evolution \cite{Jung15JPCC,Fan18SM}. Depending on the parameterization, the lithiation of silicon does not take place or is too fast as compared to experiments. In this case, interesting insights on the lithiation of $c$-SiNWs can still be gained. The Modified Embedded Atomic Method (MEAM) potential \cite{Baskes92PRB} from Cui {\it et al} \cite{Cui12JPS} has been used to understand the plasticity and toughness  of Li$_x$Si alloys with classical MD simulations \cite{Khosrownejad16JMPS, Khosrownejad17JMPS}, or with accelerated MD simulations \cite{Yan17PRM}. This potential requires much lower computational resources than ReaxFF and appears slightly better  at reproducing the elastic softening and plastic behavior of  amorphous Li$_x$Si alloys   \cite{Wang16EML}. However, the MEAM potential from Cui {\it et al.} \cite{Cui12JPS} does not succeed at modeling the lithiation of  silicon. To overcome this problem, different studies used a protocol of random lithium insertion to capture the lithiation process  \cite{Lee14MMI,Fan18SM} or the diffusive  molecular dynamics \cite{Mendez18JMPS} to accelerate the simulation and reach time scales up to  milliseconds. Nevertheless, as lithium atoms show low interactions with silicon, the results could be biased due to the poor parametrization of the chosen potential.

Here, the goal is to model the lithiation and the delithiation of the largest possible Si nanostructures and to better understand the link between lithiation, phase transformation, stress, plasticity, and fracture over several charge-discharge cycles. For that purpose, we considered the MEAM potential for its computational efficiency and  its relative simplicity to modify its parameters and  optimize its behavior. Although, they might be less robust than a ReaxFF potential  to model molecules or very exotic configurations, the MEAM potentials are usually complex enough to accurately investigate bulk, interface, and surface properties when sufficiently optimized \cite{Baskes92PRB}. 
The first part of this work will focus on optimization of the MEAM potential of Cui {\it et al.} \cite{Cui12JPS} on static properties like the energy of Li$_x$Si alloy phases and also on dynamic properties like the diffusion of Li and Si. Previous calculations of lithiation and delithiation of small $c$-SiNW from first principles methods \cite{Bunjaku} served as references. The second part of the paper will be dedicated to the lithiation and delithiation of larger amorphous Si nanowires ($a$-SiNWs) using the optimized parameters. We will then discuss the difference between crystalline and amorphous SiNWs during lithiation, the effect of temperature and finally, the history of the charge-discharge cycles on the general behavior of the Si anode, before concluding.

\section{Methods}
\subsection{Computational setup and models}

Simulations have been carried out with the open-source code LAMMPS (Large-scale Atomic/Molecular Massively Parallel Simulator) \cite{Plimpton95JCP}. Periodic boundary conditions have been applied along the three directions of space. Molecular static simulations for energy minimization have been done with the conjugate gradient algorithm until forces become smaller than $10^{-6}$ eV/\AA. Molecular dynamics simulations have been performed in the isothermal-isobaric ensemble (NPT) {\it i.e.} at constant number of particles N, constant pressure P (0 Bar) and constant temperature T, using the Nos\'e-Hoover thermostat. A time step of 1 fs assures the stability during the integration of motion equations. The damping parameter has been set to 1 ps to control the temperature and 1 ns for the pressure. The inter-atomic forces on silicon and lithium atoms have been modeled by a MEAM potential \cite{Baskes92PRB}. For pristine silicon, the initial parameters of Lee {\it et al.} \cite{Lee07CCPDT} has been used. They are known for their good mechanical properties. They were recently modified to improve the plasticity properties of silicon as well as the amorphous properties \cite{Godet16EML}. For the Li and Li-Si interactions, we have optimized the original version of Cui {\it et al.} \cite{Cui12JPS} in order to better represent the lithiation of silicon nanowires according to DFT calculations \cite{Bunjaku} (see next part). 
The atomic visualizations and parts of the different characterizations have been done with the open-source software OVITO \cite{Stukowski10MSMSE}. The {\it ab initio} calculations have been done with the Vienna Ab-Initio Simulation package (VASP) \cite{Kresse96PRB, Kresse96CMS}. The details of these calculations can be found in a previous paper \cite{Bunjaku}.

\begin{figure}
   \centering
   \includegraphics[width=8.6cm]{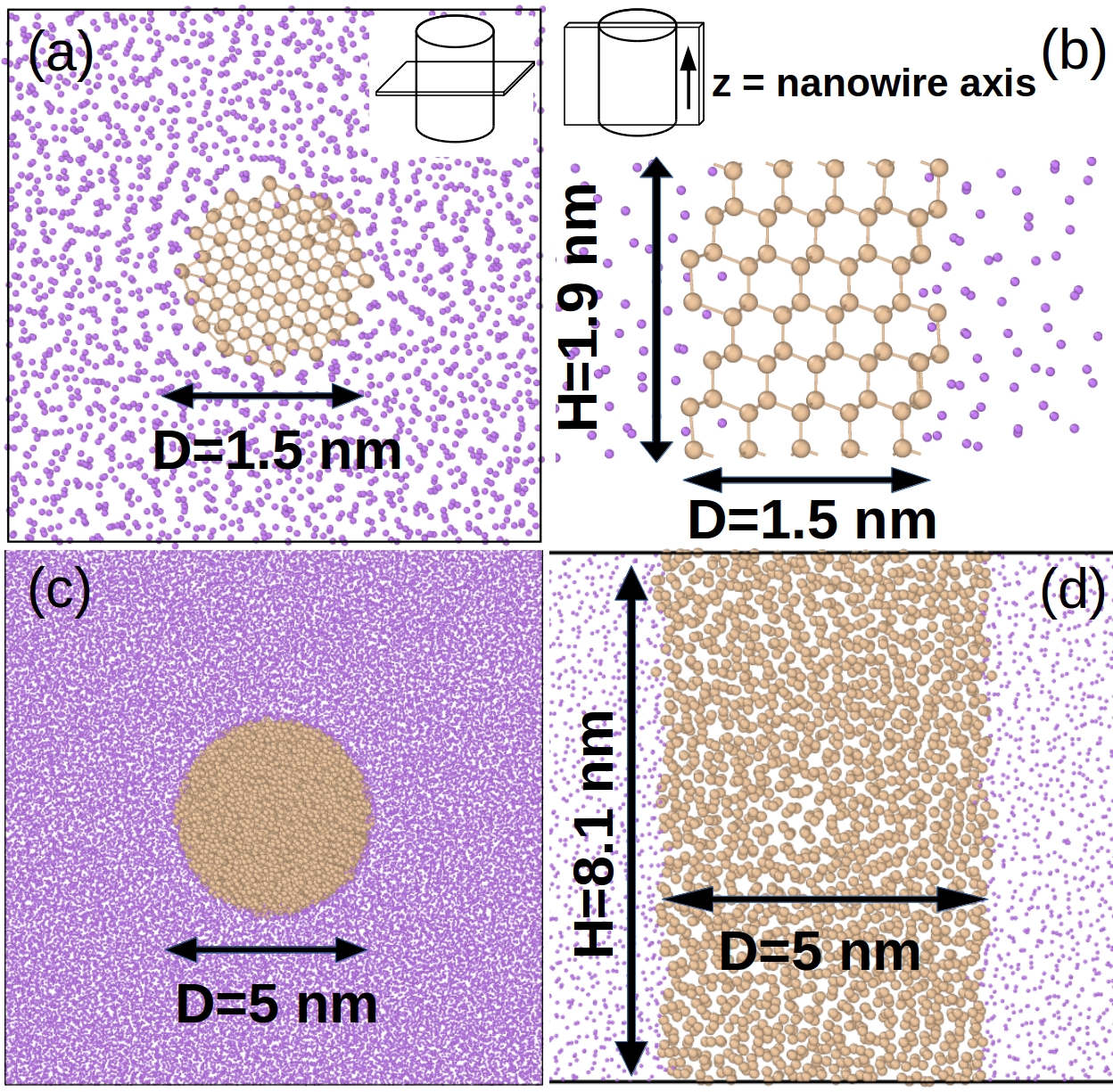}
   \caption{Model structures of $[111]$ crystalline (a-b) and amorphous (c-d) circular NWs before lithiation. Large brown spheres represent Si atoms, small violet dots Li atoms respectively. (See electronic version for color). (a and c) correspond to cross section views of crystalline and amorphous NWs, (b and d) are slices of front views of crystalline and amorphous NWs, respectively.} 
   \label{model}
\end{figure}

The considered crystalline silicon nanowires ($c$-SiNW) are cut in diamond crystal bulk with the nanowire axis orientated along the $[111]$ direction, while amorphous NWs are cut in amorphous bulk (Fig.~\ref{model}). The smallest circular NWs has a diameter of 1.5 nm  and a height of 1.9 nm, the largest a diameter of 5 nm and a height of 8.1 nm. They are first thermalized at 750K and quenched to 0K before subsequent lithiation. The amorphous bulk is obtained by creating a random structure of silicon atoms in accordance with the silicon amorphous density. A few conjugate gradient steps minimization are done to prevent overlapping atoms. Then an annealing at 4500K over 100 ps, followed by a quench to 1K over 400 ps and a final conjugate gradient minimization are performed. This method was already validated in a previous work \cite{Guenole13PRB}.

For lithiation, Si NWs are embedded in a bcc Li structure, such that all Li atoms are at a distance from a Si atom greater than the average Li-Si bonding distance of 2.7 \AA\, thus ensuring no overlapping. While Si atoms are kept frozen, an annealing of the surrounding Li atoms is performed above the melting temperatures of Li ($\sim$440K) over 100 ps until the lithium is homogeneously melted. 
The lithiation is then performed by running MD simulations on the full systems (Li+Si) at a constant temperature. To increase the speed rate of lithiation, we determined for the highest temperature that can be sustained by the systems and still produce lithiated system configurations in accordance with the {\it ab initio} results \cite{Bunjaku}.

For delithiation, a shell and a random methods have been compared. First, the extraction of the lithiated NW from the lithium phase is done by removing all Li atoms located at a distance greater than 3.7 \AA\ ($\sim$ the largest Li-Si bond length), followed by an annealing at 1150K over 500 ps. The shell method corresponds to an annealing of 100 ps after the outermost shell of Li atoms within a thickness of 0.5 \AA\ have been removed. This operation is repeated until full delithiation, as proposed in \cite{Jung15JPCC}, is achieved. For the random method, during an annealing, lithium atoms are randomly removed at a constant rate of one atom every 0.5 ps. On the largest NWs, the Li diffusion towards the outer shell was not efficient on the time scale and temperature range used. As the results from both methods used on the smallest NW were similar and in overall agreement with the DFT simulations \cite{Bunjaku}, we restricted ourselves to the random delithiation approach in this study. In addition, this technique leads to a homogeneous delithiation of the NWs, favoring a single phase process as supported by  experimental observations \cite{Cao17AMI}.

The calculation of the atomic stress given by the software LAMMPS uses the virial theorem and provides a stress tensor in units of "stress times volume". To access the local stress, a common way is to divide the stress tensor by the atomic volume, usually calculated with the Voronoi technique. As the volume of surface atoms is ill-defined, a fixed volume of $\sim$25 \AA$^3$, corresponding to the largest atomic volume over Li$_{x}$Si alloys, is considered here.  This method works properly in single crystal because the atomic volume is identical for every atom. However, in amorphous alloys, the volume strongly depends on the species and on the local configuration leading to  very noisy atomic stresses not usable for further analysis. Branicio and Srolovitz \cite{Branicio09JCP} proposed different methods of local averaging to solve this problem. The {\it atomic stress} tensor in units of stress is obtained by the weighted average of the $atomic\ stress \times volume$ divided by the weighted average of the {\it Voronoi atomic volume}. We tested such weighted functions and found that a cubic weighted function with a radius of 9 \AA\ gives relevant local stresses with an error bar smaller than 0.1 GPa at 1500K on the Li$_{15}$Si$_4$ phase compared to the overall cell stress tensor. In addition, to determine the local stress on the fly from MD simulations, the $atomic\ stress\times volume$ as well as the $Voronoi\ atomic\ volume$ have been time-averaged over 1000 configurations taken every 10 time steps to ensure converged values. Here, we only considered the local stress along the NW axis for geometry reason and because it undergoes the highest variation during the lithiation.

\subsection{LiSi - potential optimization}

Here, only the MEAM parameters involved in the Li-Si interactions have been modified. The optimization has been performed on relaxed configurations  (alloy phases, defect energies...) as it is usually done. In addition, we have also taken into account the dynamic properties of the lithiation of $[111]$ c-Si NW (Fig.~\ref{model}). In each case, we have run MD simulations from 450 to 1500K to determine the parameter set that best reproduces the lithiation observed in {\it ab initio} simulations \cite{Bunjaku}. The obtained parameters can be found in Supplementary Material.

Figure \ref{Alloys} shows the formation energy of different Li$_x$Si alloys. In addition to the usual alloy phases described experimentally \cite{Wen81JSSC}, we report other phases that appear during MD simulations of the SiNW lithiation with certain sets of parameters. The formation energy of Li$_n$Si$_m$ alloy is determined as:
$$E_f^{Li_nSi_m}=(E_{tot}^{Li_nSi_m}-n\mu_{Li}-m\mu_{Si})/m$$
where $E_{tot}^{Li_nSi_m}$ is the potential energy of the relaxed lithiated phase, $\mu_{Li}$ the atomic energy of Li in the bcc bulk phase, and  $\mu_{Si}$ the atomic energy of Si in the diamond bulk phase. Except for the L1$_2$ phase at x=3 that is not stable, our {\it ab initio} calculations show a decrease of the formation energy of Li$_x$Si alloys as the Li concentration increases until x=3.75. However, the higher Li concentrations are energetically less favorable, in agreement with a previous DFT study \cite{Gu13Nano}.  The results obtained with GGA are slightly higher by 0.1 to 0.2 eV/Si atom than in their LDA counterparts.  A similar trend is obtained with our new parameterization although the phase stability with low Li concentrations is under evaluated and that with high Li concentrations over-evaluated. Nevertheless, the phase energies are better than those given by the original version of Cui {\it et al.} \cite{Cui12JPS}.

\begin{figure}
   \centering
   \includegraphics[width=8.6cm]{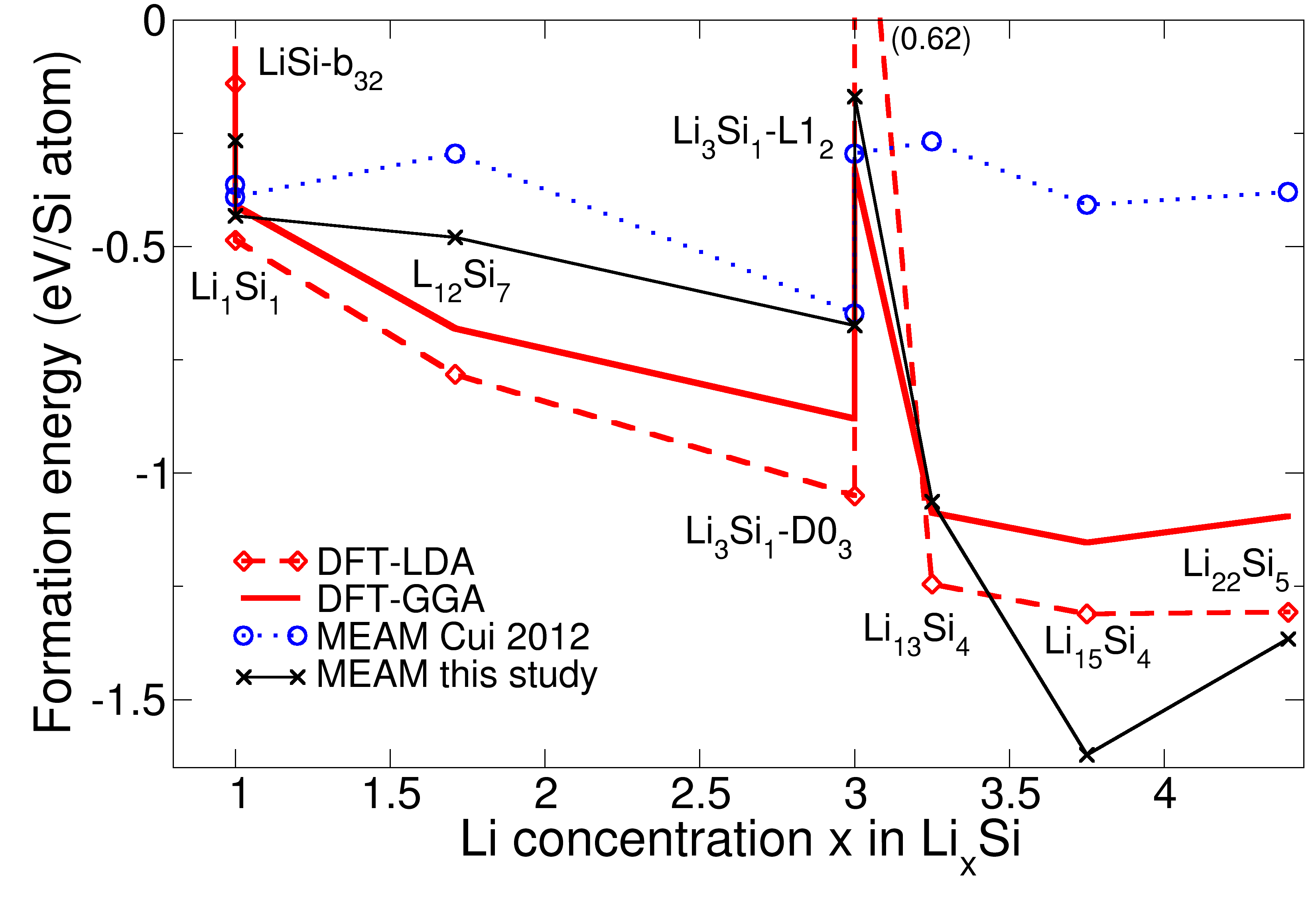}
   \caption{Formation energy of Li$_x$Si alloys, calculated with our MEAM potential (line with crosses), with DFT (LDA and GGA) and with the MEAM potential of Cui {\it et al.} \cite{Cui12JPS} (dotted line with circles). Strukturbericht designation are used to identify unusual phases.} 
   \label{Alloys}
\end{figure}

Table \ref{Edefect} summarizes the interstitial and substitutional defect energies calculated in Li and Si using the above formula, but omitting the division by the number of Si atoms $m$. Although the parameterization from Cui {et al.} and ours are not fully satisfying, we note that the energies of the interstitial Li atoms in bulk Si given by our fit are closer  to the DFT values than those of Cui. This could explain the easier diffusion of Li atoms inside silicon planes, as observed in our simulations done on $\langle111\rangle$ SiNW and presented later [Fig.~\ref{temp_on_lith_delith}(d) inset B], favoring the layer-by-layer peeling of the \{111\} planes, as expected experimentally \cite{Liu12NatNano}.

\begin{table}
\caption{\label{Edefect}Atomic defect energies in eV, for Si atoms situated in a substitutional position (Si$_{sub}$) or in an interstitial position (Si$_{int}$) in bcc Li, and similarly for Li in Si diamond bulk. Calculations were done within DFT-LDA, with the MEAM parameters of Cui {\it et al.} \cite{Cui12JPS}, and with ours. }
\begin{ruledtabular}
\begin{tabular}{lccc}
&DFT-LDA &MEAM Cui 2012&our MEAM \\
\hline
Si$_{sub}$&-0.875	&0.074	&-1.467\\
Li$_{sub}$&2.816	&0.960	&1.109\\
Si$_{int}$&-0.677	&0.298	&1.503\\
Li$_{int}$&0.048	&1.300	&0.591\\
\end{tabular}
\end{ruledtabular}
\end{table}

\begin{figure}
   \centering
   \includegraphics[width=8.6cm]{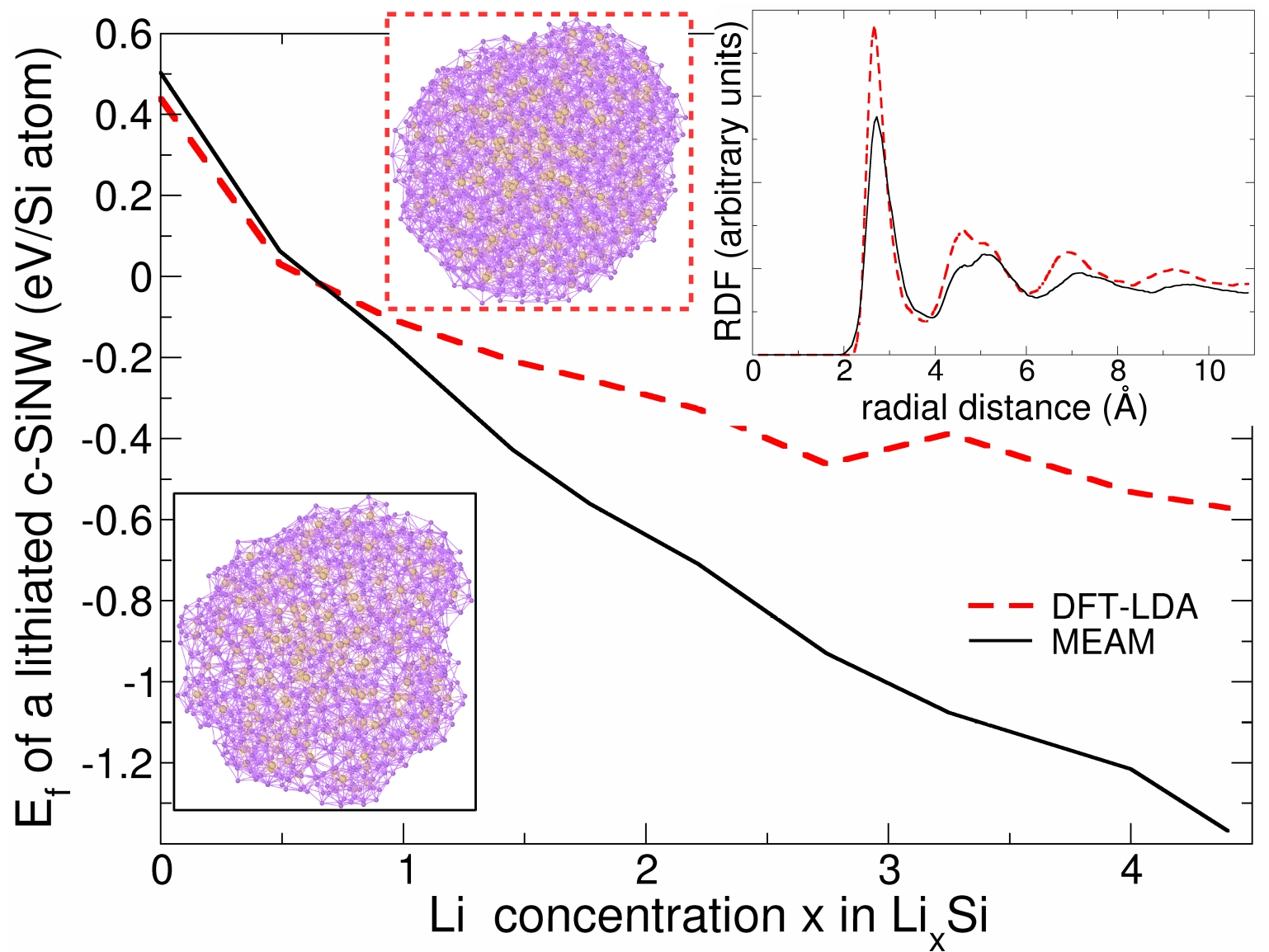}
   \caption{Formation energy (E$_f$) of a $c$-SiNW (diameter = 1.5 nm) lithiated at different Li concentration. The calculations were performed within DFT-LDA (dashed lines) \cite{Bunjaku} and with our MEAM potential (solid lines). The inset represents the Radial Distribution Function (RDF) calculated on the periodic NW obtained in DFT and with our MEAM potential. The atomic structures of the fully lithiated NWs are also depicted (DFT on the upper side, MEAM on the lower side).} 
   \label{RDF_Ef}
\end{figure}

Figure \ref{RDF_Ef} shows how our parameterization models the energy and structure of a lithiated $c$-SiNW. The relaxed configurations of a lithiated $c$-SiNW previously obtained in DFT-LDA \cite{Bunjaku}  have been annealed in classical MD simulations with our version of MEAM, during 200 ps at 750K, quenched to 0.1K in 200 ps, and followed by an energy minimization. The formation energies have been calculated with the previous equation used for the alloy phase energies. At the most lithiated stage, the radial distribution function (RDF) of the lithiated NW obtained in DFT is characteristic of an amorphous phase. The formation energy converges toward 0.6 eV/Si atom, well above the alloy phase energy with a similar Li concentration. This result is in agreement with previous {\it ab initio} studies \cite{Gu13Nano,Cubuk14NL}. It underlines the strong influence of the atomic structure and so the electronic bonding on the stability of the alloy. While the formation energies of slightly lithiated NW are similar when calculated with the MEAM potential and with DFT, the values from MEAM differ from those in DFT at higher Li concentrations, where they reach the formation energy of the most lithiated crystalline phase obtained with MEAM ($\sim$ -1.5 eV/Si atom in Fig.~\ref{Alloys}). This pinpoints the lake of transferability of empirical potentials, in particular when the electronic bonding is strongly perturbed. Note that a poor description of the surface energy by the MEAM potential could affect the formation energy at so small size where the surface-to-volume ratio  is important. However, the MEAM potential is able to keep the amorphous structure of the fully lithiated SiNW after annealing, as demonstrated by the similar RDF calculated from MEAM and DFT configurations (inset in Fig.~\ref{RDF_Ef}).

\begin{figure}
   \centering
   \includegraphics[width=8.6cm]{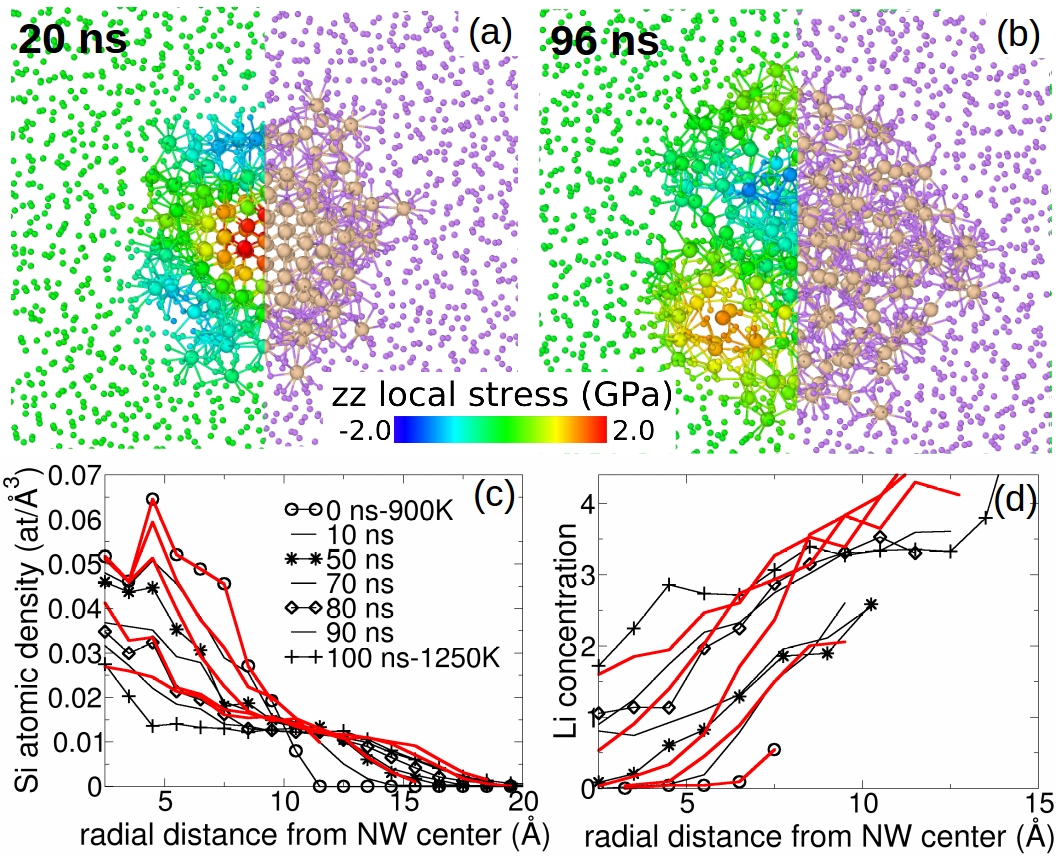}
   \caption{Lithiation of a 1.5 nm diameter $c$-SiNW (Fig.~\ref{model} top panels) with our MEAM potential after (a) 20 and (b) 96 ns. Only Si-Si and Li-Si bonds are drawn for clarity. The color map on the left part of each NWs represents the local axial stress. (c) Radial Si atomic density and (d) radial Li concentration $x$ in Li$_x$Si obtained from DFT (red bold curve) and with the MEAM potential (black thin curves) at different lithiation stages.} 
   \label{lith_15Ang}
\end{figure}

Figure \ref{lith_15Ang} compares the different states of lithiation on small $c$-SiNW (Fig.~\ref{model} top panels) obtained from first-principles \cite{Bunjaku} and with our MEAM potential. Over 100 ns, a linear temperature ramp from 900 to 1250K is applied in the MD simulations. Even at this small scale, the two-phase lithiation mechanism is clearly observable [Fig.~\ref{lith_15Ang}(a)], with an amorphous Li$_x$Si alloy shell growing on a pristine Si core, in agreement with {\it ab initio} \cite{Bunjaku} and experimental results \cite{Liu12NatNano}. Despite the very small dimensions of the system, the lithiation time is very prohibitive to reach a fully lithiated NW (96 ns). During the two-phase lithiation mechanism, the core of the SiNW experiences a tensile stress of about 2 GPa along its axis, as revealed by the zz local stress map [Fig.~\ref{lith_15Ang}(a)]. The stress is relaxed after the NW is fully lithiated [Fig.~\ref{lith_15Ang}(b)]. The kinetics of the lithiation observed in MD simulations have been compared to DFT results \cite{Bunjaku} through the radial Si atomic density and Li concentration evolution over the full lithiation stages [Fig.~\ref{lith_15Ang}(c) and  \ref{lith_15Ang}(d)]. The radial densities are evaluated on cylindrical shells with a thickness of 1 \AA. Each curve results from a running averaged over 4 \AA. The Si density  as well as the Li concentration profiles are relatively similar between DFT and MEAM, which confirms the ability of our potential to reproduce the lithiation on small systems. At the fully lithiated state, the Si density reaches a constant value through the NW close to 0.013 atom/\AA$^3$. The Li concentration also reaches a plateau, at least for the MEAM potential, with a concentration of 3.25 Li atoms per Si atom, slightly lower than the expected experimental value of 3.75 \cite{Gu13Nano}. However, as in experiments \cite{Gu13Nano} no intermediate crystalline Li$_x$Si alloy phase is observed. 

A similar study has also been done for the delithiation of small $c$-SiNW (not shown here). The delithiation procedure with the MEAM potential reproduces relatively well the trends obtained in DFT \cite{Bunjaku}. More details will be discussed on the delithiation of larger NWs here-after. 
We are now confident that our potential is robust and reliable enough to correctly model the lithiation/delithiation cycle on small $c$-SiNW. We will use it in the next section to investigate larger NWs, in particular those with an amorphous structure.

\section{Lithiation and delithiation of $a$-SiNW}

\subsection{Lithiation of amorphous SiNW}

\label{sec_a-lithiation}

\begin{figure}
   \centering
   \includegraphics[width=8.6cm]{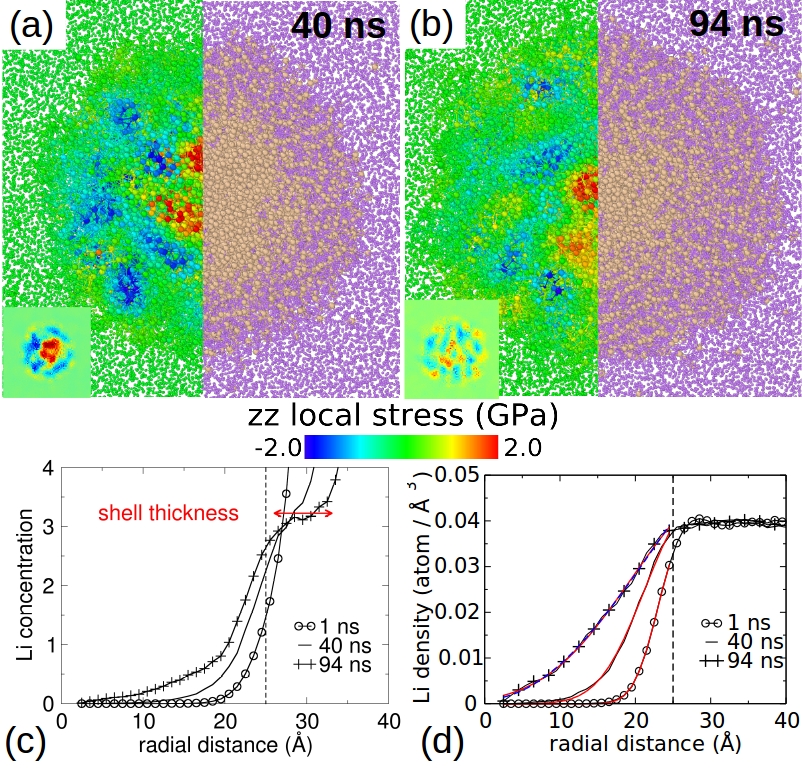}
   \caption{Lithiation of an amorphous SiNW (diameter$\sim$5 nm, height$\sim$8 nm) at 1150K. (a-b) Cross section views of the atomic configurations with their local axial stress map at t=40 and 94 ns, respectively. The insets are the axial average of the local stress. (c) Radial Li concentration ($x$ in  Li$_x$Si alloy) and (d) radial Li density  as a function of the radial distance. The vertical dotted lines are the radius of the $a$-Si NW before lithiation. The full red and dashed blue curves have been fitted on diffusion equation solutions (see text). } 
   \label{a-Si_lithiation}
\end{figure}

The long simulations times ($\sim$100 ns) required for the lithiation of a small $c$-SiNW, as shown in the previous section (Fig.~\ref{lith_15Ang}), prevent the investigation of very large systems. A consequence, the largest system that has been considered has a diameter of 5 nm and a height of 8 nm (Fig.~\ref{model}). The lithiation over  100 ns already represents a cost of 60000 CPU hours. After 94 ns, the lithiation results in an amorphous shell of Li$_x$Si alloy [Fig.~\ref{a-Si_lithiation}(a) and \ref{a-Si_lithiation}(b)], with a thickness of about 8 \AA\ and a lithium concentration $x$=3.25 [Fig.~\ref{a-Si_lithiation}(c)], equal to the maximum Li concentration observed with this potential during lithiation. The corresponding shell growth speed is about 0.08 \AA/ns, at least 4 orders of magnitude faster than the experimental speed observed in amorphous Si nano-spheres \cite{McDowell13NL}. While several theoretical works assume that the lithiation of $a$-Si is always a diffusion-controlled single-phase process \cite{Mendez18JMPS}, experimentally, on $a$-Si nano-spheres \cite{McDowell13NL} and on planar interface \cite{Wang13NL}, a sharp concentration gradient during the first lithiation is observed, suggesting a two-phase lithiation mechanism. Here, our results effectively support the experimental observation of a two-phase lithiation mechanism, as in $a$-Si nanoclusters in DFT \cite{Pedersen17PRA}, with a  fully lithiated shell and an $a$-SiNWs core. They also suggest a chemical diffusive process between both phases on a relatively wide interface. The complexity of the Li diffusion in $a$-Si related to a large interface diffusion and a two-phase lithiation mechanism could explain why some experimental works proposed a single phase diffusion \cite{Cao17AMI}, while others a two-phase diffusion \cite{McDowell13NL,Wang13NL}. It is worth noting that the accelerated diffusive molecular dynamics is not able to capture this lithiation mechanism in $a$-Si \cite{Mendez18JMPS}. Here, our results underline the limit of the method for studying local atomic mechanisms.

The analysis of the Li diffusion  between the $a$-SiNW core and the fully lithiated shell is shown in Fig.~\ref{a-Si_lithiation}(d). To extract the diffusion coefficient of Li in the $a$-SiNW, we considered the solutions of Fick's equation in an infinite cylinder with a constant Li concentration on its surface and a diffusion coefficient independent of the Li concentration \cite{Crank75}.  For the maximum lithiated configuration (t=94 ns) the lithium atoms almost reached the center of the NW so that a solution  based on Bessel functions can be applied [Fig.~\ref{a-Si_lithiation}(d) blue dashed curve]. For the weakly lithiated configurations (t=1 and 40 ns), the Li atoms did not reach the center of the NW and the previous solution does not work. We then considered an approximative solution based on an error function, as proposed by Crank \cite{Crank75}  [Fig.~\ref{a-Si_lithiation}(d) red curves]. The fitted curves superimpose very well on the Li density curves extracted from our MD simulations, underlying the relevance of the models. The correlation parameters are higher than 0.999 for the three curves. The diffusion coefficient is $13.7 \times 10^{-7}$ cm$^2$s$^{-1}$, $1.15 \times 10^{-7}$ cm$^2$s$^{-1}$, and  $0.68\times 10^{-7}$ cm$^2$s$^{-1}$ for t=1, 40, and 94 ns, respectively. The approximative solution applied to the state t=94 ns leads to  $1.8 \times 10^{-7}$ cm$^2$s$^{-1}$  close to the result given by the exact solution method. The diffusion coefficients obtained here at 1150K are  2 order of magnitude higher than the fastest values found in  literature for single crystalline NW at 700K \cite{Seo15RSCA}, but close to bulk alloy values at 700K \cite{Wen81JSSC}.  The high temperature and the amorphous phase considered in our simulations could effectively explain the large diffusion coefficients. The small size of our NWs could also plays a role because the diffusion coefficient decreases as the lithiation goes on. As the fitting curves based on a diffusion coefficient independent of the Li concentration are in excellent agreement with our data, the variation of the coefficients over time is probably more related to the presence of a fully lithiated shell that prevents the diffusion of Li inside the Si core, rather than to the variation of the Li concentration between the alloy shell and the SiNW core. This assumption is confirmed by the analysis of the lithiation front speed at half height on the Li density curves. It starts at 2.2 \AA/ns at 1 ns before decreasing to 0.07 \AA/ns between 40 and 94 ns. This shows the higher reactivity of Li atoms with Si at the onset of the lithiation than after a while. The lithiation appears to be more limited  by Li diffusion in the Li$_x$Si alloy than by the interface reaction, as assumed experimentally \cite{McDowell13NL}. 

In $a$-SiNW, the stress maps are relatively inhomogeneous along the radial and the axial directions [Fig.~\ref{a-Si_lithiation}-(a) and \ref{a-Si_lithiation}(b)]. For a better insight into the stress distribution of the lithiated $a$-SiNW, we can consider its axial average [insets in Fig.~\ref{a-Si_lithiation}(a) and \ref{a-Si_lithiation}(b)]. The first stage of  lithiation (t=40 ns) of the $a$-SiNW is similar to the lithiation of $c$-SiNWs: the fully lithiated outer shell appears almost free of stress, the partially lithiated intermediate shell with a compressive stress  and the pristine Si core with a tensile stress. Note that a stress of about 2 GPa is in very good agreement with the experimental value on amorphous thin film \cite{Sethuraman10JPS}.  After a longer lithiation (t=94 ns), the local and the average stresses are still inhomogeneous without any evidence of core-shell stress configurations, with slightly lower stress intensity compared to the case at t=40 ns. The Li diffusion over large distance toward the NW center could then form Li$_x$Si alloys on thick interface.  Conversely to pristine Si, these alloys  are known to undergo plastic deformations at low stress to relax the stored energy \cite{Zhao11NL, Fan13MSMSE, Wang16EML, Khosrownejad16JMPS}.  The inhomogeneity of the stress could be a result of the localized plasticity in the amorphous Li$_x$Si alloys as previously observed by Yan {\it et al.} \cite{Yan17PRM}.

\subsection{Delithiation of amorphous NW}
\label{delithiation}

\begin{figure}
   \centering
   \includegraphics[width=8.6cm]{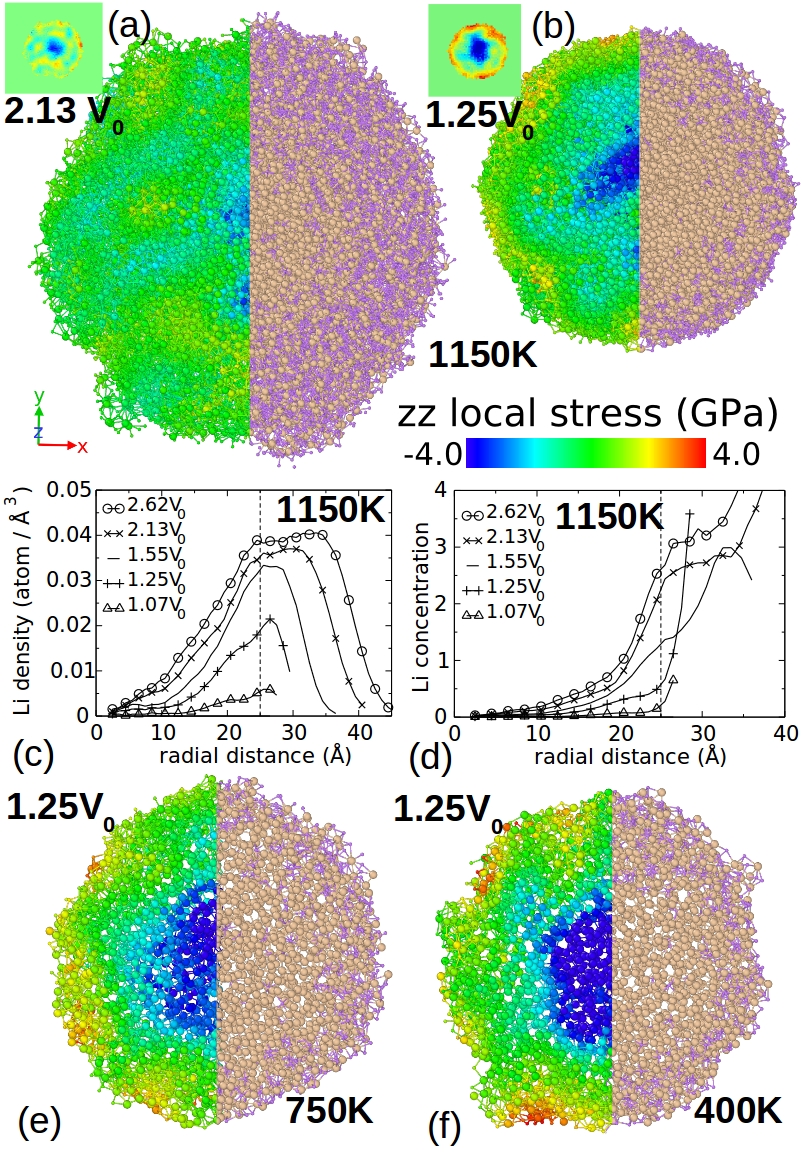}
   \caption{Delithiation of amorphous NWs  (diameter$\sim$5 nm, height$\sim$8 nm) at 1150K. (a-b) Atomic configurations with stress map and axial averaging stress in inset. (c) Li density and (d) Li concentration $x$ in Li$_x$Si as a function of the radial distance at different time. The vertical dotted lines are the radius of the $a$-Si NW before lithiation. (e-f) Slices of delithiated NWs at 750 and 400K, respectively. } 
  \label{a-Si_delith}
\end{figure}

The  expanded volume of the most lithiated configuration (13626 Li atoms for 7955 Si)  is about 2.62$V_0$, $V_0$ being the initial volume of the pristine $a$-SiNW, in the range of the experimental values observed for $a$-Si nanospheres \cite{McDowell13NL}. Figure \ref{a-Si_delith}(a-d) shows the delithiation at 1150K. From 2.62 to 1.55$V_0$  [Fig.~\ref{a-Si_delith}(c)], the density of Li atoms remains almost constant, whereas the radial distribution of Li is noticeably reduced. During delithiation, the system evolves to keep a high  Li concentration close to x=3.25 by reducing the thickness of the Li$_x$Si alloy shell, as shown by the Li concentration curves [Fig.~\ref{a-Si_delith}-(d)]. In addition, for the advance delithiation stages (1.25 to 1.07$V_0$), the lithium atoms are more located on the surface, as observed in DFT \cite{Bunjaku}. 
At the fully delithiated stage, and independently of the temperature (1, 400, 750, or 1150K), the NW is still amorphous with the same volume $V_0$ as before  lithiation. The $a$-SiNW is then completely healed after fully removing the Li atoms even at very low temperatures. It appears that as soon as Li atoms are present in the structure, they greatly improve the plasticity of the amorphous, in agreement with DFT results \cite{Zhao11NL}. The experimentally observed 25\% volume increase at the end of the delithiation of $a$-Si nano-spheres \cite{McDowell13NL} is probably due to the presence of remaining Li atoms rather than to  a porous amorphous silicon. Similar results have been found with DFT \cite{Bunjaku}.
 
During the delithiation at 1150K, a compressive stress is progressively built in the center of the NW, reaching -5 GPa at 1.25$V_0$, while the outer shell reaches a  tensile stress of 3.1 GPa [Fig.~\ref{a-Si_delith}(a) and \ref{a-Si_delith}(b)]. Here again, the stress concentrations manifest themselves where the Li density is low, giving a less efficient stress relaxation through  plasticity. As it is well known in amorphous materials, the plastic flow for stress relaxation decreases at low temperatures; here also,  when the temperature is 750 and 400K, the magnitude of the inner stress reaches -7.6 GPa and -9.4 GPa, while in the outer stress it reaches 3.8 and 4.2 GPa, respectively [Fig.~\ref{a-Si_delith}(e) and \ref{a-Si_delith}(f)].  This study underlines the relatively strong stress concentration during delithiation, approaching the theoretical yield stress of crystalline silicon nanopillars ($\sim$10 GPa) \cite{Merabet18AM}, when the temperature is close to the working condition of the cell (300K). Interestingly, the larger stress magnitudes observed during delithiation compared to lithiation agree with the asymmetrical stress observed experimentally  in crystalline silicon \cite{Chon11PRL}.

\section{Discussion}

\subsection{Difference in stress release during the lithiation of $a$-SiNW and $c$-SiNW}

\begin{figure}
   \includegraphics[width=8.6cm]{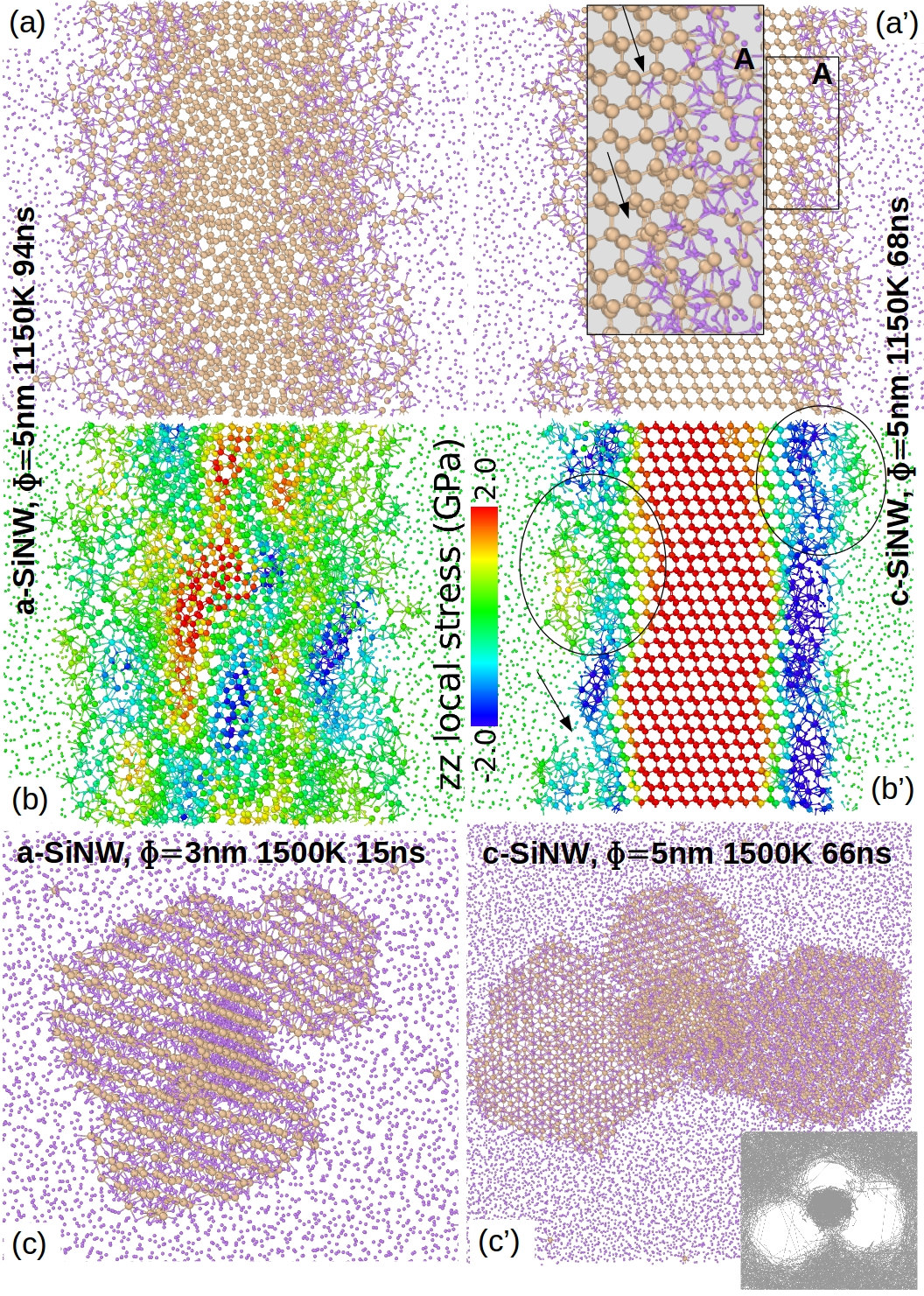}
   \caption{Slice views of lithiated (a) $a$-SiNW  and (a') $c$-SiNW  with similar dimensions (diameter$\sim$5 nm, height$\sim$8 nm) at 1150K, with their respective axial local stress map in (b) and (b'). The insets are zooms on atomic structures. Cross section of  lithiated (c) $a$-SiNW and (c') $c$-SiNW at high temperature (1500K).  The high temperature accelerates Li diffusion and leads to crystalline grain growth around the core of the NW. The inset in (c') represents the trajectories of Li atoms during the last 30 ns of lithiation.} 
   \label{temp_on_lith_delith}
\end{figure}

The large diffusion interface observed during the lithiation of $a$-SiNW described in section \ref{sec_a-lithiation}  differs from the two-phase lithiation mechanism found in $c$-SiNW [Fig.~\ref{temp_on_lith_delith}(a) and \ref{temp_on_lith_delith}(a')]. In $c$-SiNW, it appears that the lithium atoms diffuse between the (111) Si atomic planes on a short distance, until reaching a high enough Li concentration and consequently breaking the Si-Si bonds to form an amorphous phase (see inset A). The interface is limited to  $\sim$1 nm, in agreement with the  atomic peeling mechanism described experimentally \cite{Liu12NatNano}, and corresponds to the Si-Si bonds breaking thanks to charge transfer from lithium to silicon atom, as detailed in DFT \cite{Bunjaku}. 

While the stress map on lithiated amorphous NW [Fig.~\ref{temp_on_lith_delith}(b)] exhibits almost no pattern, as discussed in section \ref{sec_a-lithiation} with stress magnitude in between -2.4 and 2.9 GPa, the lithiation of $c$-SiNW presents a core-multi-shell pattern. In $c$-SiNW our calculations lead to the formation of a compressive  Li$_x$Si alloy outer shell, as observed experimentally \cite{Liu13Nano} and theoretically \cite{Jung15JPCC,Fan18SM, Mendez18JMPS}. More precisely, our results show that the crystalline Si core is under an axial tensile stress of up to 3.5 GPa, and only the thin interface presents a relative important compressive axial stress that can reach $-$3.5 GPa [Fig.~\ref{temp_on_lith_delith}(b')]. The outermost shell formed of amorphous Li$_x$Si alloy is however almost unconstrained which differs from previous works  \cite{Liu13Nano, Jung15JPCC,Fan18SM, Mendez18JMPS}.  The transition from the Li-rich cubic diamond Si phase to the amorphous Li$_x$Si alloy appears as a none continuous mechanism, and takes the form of a localized  spontaneous phase transition that releases the interface stress [left ellipse in Fig.~\ref{temp_on_lith_delith}(b')]. When the lithiation is pursued, the accumulation of Li atoms between the amorphous alloy and the Si core generates again a compressive stress at the interface [right ellipse in Fig.\ref{temp_on_lith_delith}(b')]. For almost the same lithiation duration, the crystalline SiNW is  weakly  lithiated compared to the amorphous one. As in both cases the outermost shells are formed of a similar alloy, the retardation of lithiation should be controlled by the low Li diffusion at the interface of $c$-SiNW. This can be attributed to the compressive stress generated under lithiation, as also observed with the ReaxFF potential \cite{Ding17NE,Ostadhossein15PCCP} and proposed experimentally to explain the self limiting lithiation of (111) SiNW \cite{Liu13Nano}. Conversely to the lithiated amorphous structures that behaves like glass materials with easy plastic flow to relax the stress, the crystalline interfaces prevent any stress release and progressively increase the  diffusion barrier for the Li atoms as the interfaces become highly lithiated.

This mechanism of 'spontaneous' phase transition from crystalline to amorphous alloy is at the origin of the wavy structure observed on long NWs [Fig.\ref{temp_on_lith_delith}(b')]. These small perturbations could lead to the inhomogeneous lithiations observed experimentally over multiple lithiation - delithiation cycles \cite{Liu12NatNano,Lee12PNAS}, or to the initiation of partially detached alloy nanograins that could induce cracks at full lithiation [arrow in Fig.~\ref{temp_on_lith_delith}(b')] \cite{Rhodes10ECS}. In addition, the stress variation from -3.5 GPa to 3.5 GPa on a narrow interface between the crystalline Si core and the alloy shell (about 5 atomic layers) generates very intense local stress gradient compared to the $a$-SiNW which could also explain why  $c$-SiNWs are more sensitive to crack formation and then less suitable for battery cells than $a$-SiNWs.

\subsection{Effects of temperature and highly lithiated structures} 

A high temperature of 1500K has been tested to accelerate the Li diffusion in $a$-SiNW and $c$-SiNW [Fig.~\ref{temp_on_lith_delith}(c) and \ref{temp_on_lith_delith}(c')]. For both NWs, fast lithiation allows to reach a fully lithiated level of the NWs. This also leads to the growth of several crystalline Li$_x$Si alloy grains around the Si core with a lithium concentration of about $x$=3.25. Interestingly, the crystallization of the fully lithiated Si structures has been reported with a ReaxFF potential at 1200K \cite{Ostadhossein15PCCP}, but also experimentally when a too low voltage is applied for the lithiation or at too high temperature or for a too long lithiation   \cite{Hatchard04JECS, Obrovac04ESSL}. Note that the doping can also strongly influence the lithiation front speed as well as the crystallization \cite{Liu11NL_Zhang}. Finally, other studies  report a spontaneous transformation from amorphous to crystalline  solely controlled by the lithium concentration in amorphous Li$_x$Si when $x$=3.75 \cite{Gu13Nano}. 

During lithiation the grains grow with different orientations around the SiNW core and form grain boundaries in between. The radial expansion is then at the origin of grain separation along  boundaries [Fig.\ref{temp_on_lith_delith}(c) and \ref{temp_on_lith_delith}(c')], with configurations that are similar to those reported experimentally on largely lithiated (111) SiNWs \cite{Lee12PNAS}. This division of material could be at the origin of the cracks observed experimentally on Si nano-particles during the first lithiation cycle  \cite{Rhodes10ECS}. In the smaller NWs (not shown here), similar simulations lead to the formation of a unique crystal grain without crack or material division. This size effect on the crack formation of highly lithiated $c$-SiNW and $a$-SiNW with large diameters was also observed experimentally at much larger scales  \cite{Liu12ACSNano}. The difference in diameter for the transition is probably due to the high temperature required in simulation to get fully lithiated structures on reachable simulation times. Finally, the lithium diffusion paths recorded during the last 30 ns preceding the  snapshot in Fig.~\ref{temp_on_lith_delith}(c') are shown in inset. They reveal that diffusion paths are only located at the grain boundaries which could help the detachment of the grains from the SiNW core.

\subsection{Cyclic lithiation and delithiation of $a$-SiNW}

Experimentally, the delithiation of $a$-Si nanospheres shows a final volume 25\% larger than the initial volume of non lithiated particles \cite{McDowell13NL}. As our fully delithiated structures present the same volume as before lithiation (see section \ref{delithiation}), we concluded that lithium atoms should have stayed trapped in Si after delithiation. To account for this effect, we stopped the delithiation when the volume of the NW reaches 1.25$V_0$, before starting the second lithiation cycle. 

\begin{figure}
   \centering
   \includegraphics[width=8.6cm]{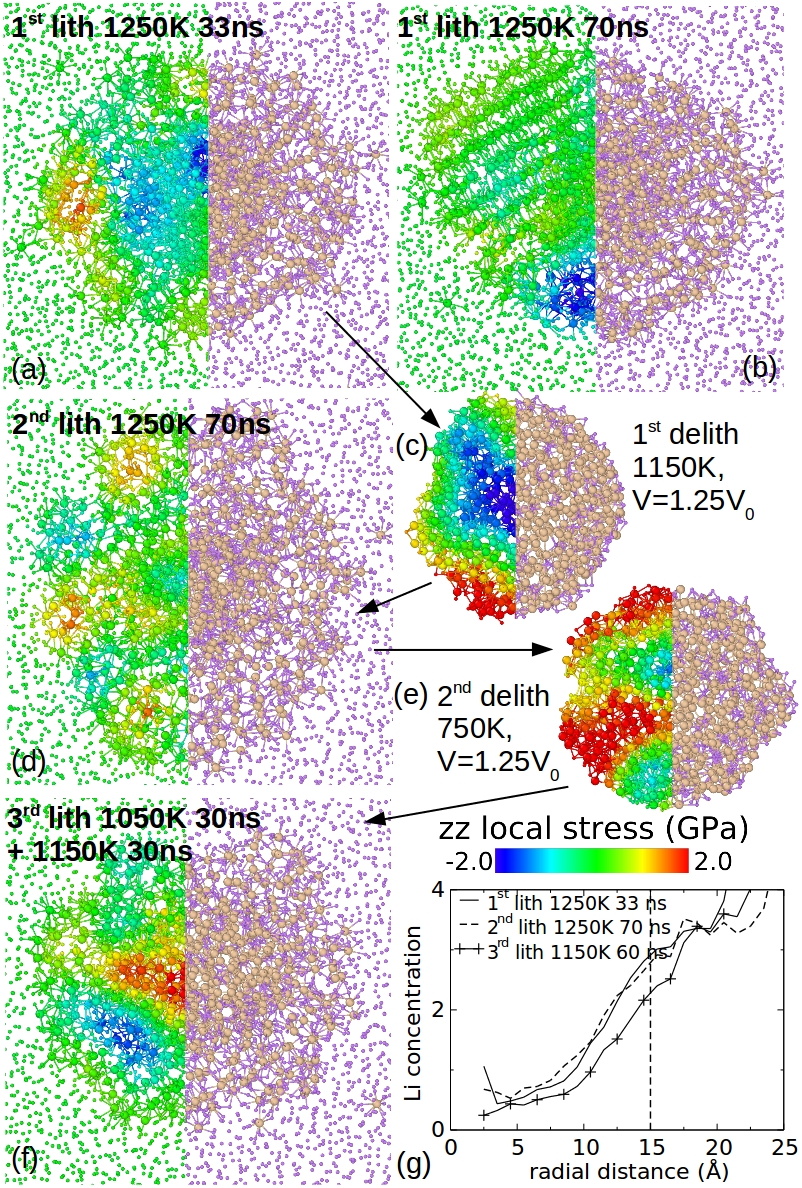}
   \caption{Cycles of lithiation-delithiation of $a$-SiNW with a diameter of 3 nm, Atomic structures with their local axial stress have been represented.  (a-b) First lithiation at 1250K after 33 and 70 ns, respectively. (c) First delithiation at 1150K. (d) Second lithiation at 1250K over 70 ns. (e) Second delithiation at 750K. (f) Third lithiation at 1050K during 30 ns then at 1150K for 30 ns. (g) Li concentration $x$ in Li$_x$Si versus radial distance after each lithiation.}
   \label{cycles}
\end{figure}

The first cycle of lithiation and delithiation has been already described in sections \ref{sec_a-lithiation} and \ref{delithiation}. Here, the lithiation has been done at slightly higher temperatures than in the previous sections to accelerate the diffusion. However, a crystalline phase appeared after 33 ns [Fig.\ref{cycles} (a-b)]. In order to perform lithiation/delithiation cycles on amorphous phases only,  as in experiments  \cite{McDowell13NL,Cao17AMI}, we used the  amorphous configuration  obtained at 33 ns for the delithiation stage (Fig.~\ref{cycles}). 
The 2$^{nd}$ lithiation is very similar to the first one [Fig.~\ref{cycles}(d) and (g)], but in this case we did not observe the crystallization for long lithiation which depicts the spontaneous and versatile character of this phase transition, as discussed in the previous section. As a consequence of the longer lithiation, the amorphous shell of the Li$_x$Si alloy becomes thicker compared to that over the first lithiation.
The 2$^{nd}$ delithiation and the 3$^{rd}$ lithiation give rise to the crystallization of the alloy. Hence, we had to adapt our simulation setup of delithiation and lithiation by decreasing the temperature, or using increasing temperature plateaus to favor the lithium diffusion without the formation of  crystalline phases. 
As the Li atom concentration remains similar over the cycles of lithiation and delithiation, we do not expect that the presence of Li atoms left from the precedent lithiation accelerate the Li diffusion. The stress maps are also characteristic of the lithiation and delithiation of $a$-SiNW. The slight modifications over cycles could be more related to the variation of temperature used or to the level of lithiation reached.  However, it appears that the history of the charge/discharge leaving Li atoms in the cell could be at the origin of the crystallization during subsequent cycles, which is known to be detrimental for the cycle performance of battery cell \cite{Obrovac04ESSL}. The formation of crystalline grains exposed to large tensile and compressive stresses during the lithiation/delithiation cycles could initiate crack formation  in $a$-SiNW anodes, reinforcing the history effect observed on void nucleation in amorphous alloy \cite{Khosrownejad17JMPS}.
In addition, the consecutive cycles decrease the temperature at which the crystallization appears. This could be understood as the aging of glassy alloy through the competition between time scale of charging/discharging and that of glassy relaxation, as proposed by Khosrownejad {\it et al.} \cite{Khosrownejad16JMPS}.

\section{Conclusion}

In this work, we first optimized a MEAM potential to more accurately model the Li$_x$Si alloy phases and the lithiation of small $c$-SiNWs based on DFT calculations. Thanks to this new potential parameterization, lithiation and delithiation of $a$-SiNWs have been realized and compared to experiments and previous theoretical works. Our results confirm that the first lithiation of $a$-SiNWs is effectively a two-phase mechanism with a fully lithiated outer shell of amorphous Li$_x$Si alloy and a pristine $a$-Si core. In addition, the calculations reveal that lithium atoms diffuse between the two phases over a much larger interface than in $c$-SiNWs. The interface thickness increases as the lithiation progresses. Inside this interface, the diffusion coefficient of Li is estimated at $0.68\times 10^{-7}$ cm$^2$s$^{-1}$, which is two orders of magnitude larger than reported experimentally for crystalline $c$-SiNWs  \cite{Seo15RSCA}, probably due to the high temperatures, the amorphous structure, and the still too small dimensions of NWs used in our study. 

As for $c$-SiNWs, the lithiation of $a$-SiNWs  overall leads to compressive axial stress in the outer lithiated shell. The lower stress magnitude observed in $a$-SiNWs compared to $c$-SiNWs has been attributed to glassy plasticity relaxation in the soft amorphous Li$_x$Si alloy phases. The effect of temperature as well as the effect of charge/discharge cycle has been investigated. Both effects result in phase transformations from amorphous to crystalline alloys. In particular, the crystallization appears at lower temperatures as the number of cycles increases. This effect is attributed to the smooth plastic glassy behavior that operates during the lithiation and also to the partial delithiation that both help the system to reach a more stable state of crystalline phase.

The size and geometry of the anode could favor the growth of grains with different orientations.  The separation of these grains along their boundaries under the large tensile and compressive stresses generated during the charge/discharge could lead to the fracture or to the material division observed experimentally. In addition, the diffusion of Li atoms inside the grain boundaries appears to help the grains separation. 

As our fully delithiated $a$-SiNWs are very similar to the original NWs before lithiation, we think that the history of cycles could be removed by full delithiations. This conclusion supports the experimental observation that the predominant failure mechanism of the silicon electrode is related to incomplete
delithiation of the silicon electrode during cycles \cite{Yoon15JECS}.
Our study suggests that more efforts should be done to improve the delithiation of silicon anodes in order to increase the cycle performance of Si-based lithium-ion battery cells.

\section{Acknowledgments}

JG would like to thank ETH Zurich for the visiting Professor position that he got from September 2018 to June 2019 at the Integrated Systems Laboratory in the Department of Electrical Engineering and Information Technology. 
Computations have been performed on the supercomputer facilities of the M\'esocentre de calcul Poitou-Charentes.


\bibliographystyle{apsrev4-1.bst}
\bibliography{biblio.bib}

\end{document}